\documentstyle[aps,twocolumn,maps]{revtex}
\begin{document}
\title{Neutrino oscillations in curved spacetime: 
	an heuristic treatment}
\draft
\author{Christian Y. Cardall and George M. Fuller}
\address{Department of Physics, University of California, San Diego,
La Jolla, California 92093-0319}
\date{October 25, 1996}
\mabstract{
We discuss neutrino oscillations in curved spacetime. 
Our heuristic approach can accomodate matter effects and
gravitational contributions to neutrino spin precession 
in the presence
of a magnetic field.
By way of illustration, we perform explicit calculations
in the Schwarzschild geometry. In this case, gravitational
effects on neutrino oscillations are intimately 
related to the
redshift.
We discuss how spacetime curvature could affect the
resonance position and adiabaticity of matter-enhanced
neutrino flavor conversion. \\ \\
PACS number(s): 14.60.Pq, 95.30.Sf, 26.30.+k}

\mmaketitle
\narrowtext
\section{Introduction}
Two decades ago an experimental connection between quantum
mechanics and gravitation was observed \cite{colella}---essentially,
a gravitational analog of the Aharanov-Bohm effect. These
neutron interferometry experiments can be  well described
by including a gravitational potential energy in the Hamiltonian
of the nonrelativistic Schr\"odinger equation.
More recently, 
the effects of gravitation on another quantum mechanical 
phenomenon---neutrino oscillations---have been
discussed by a number of authors \cite{ahluwalia,grossman,BHM,PRW}, 
some of whose results appear to 
conflict. In this paper we will provide a simple 
framework for studying
neutrino oscillations in curved spacetime. We hope to clarify some
issues left unclear by previous treatments 
\cite{ahluwalia,grossman,BHM}, 
and provide a more transparent route to some of the 
results obtained in
Ref. \cite{PRW} by more formal methods than we employ here.

As an illustration of our treatment, we will do explicit
calculations in  
Schwarzschild geometry. For radially 
propagating neutrinos, the oscillation formulae will be different
from those that appear in flat spacetime. This is not surprising,
since the gravitational redshift for radial 
neutrino propagation is well known \cite{fuller96}. 
The existence of
a circular orbit for massless particles in the Schwarzschild geometry
allows computation of the oscillation phase for 
purely azimuthal propagation
as well. In this case, it will be seen that the oscillation 
formulae are
identical to those obtained in flat spacetime.

We hope our calculations will clarify 
some issues that we feel have
been left unclear by previous treatments. 
The authors of Refs. \cite{ahluwalia,BHM} (who in turn follow Ref. 
\cite{stodolsky}) 
use a semiclassical approximation
that employs the action of a massive particle as a quantum phase
for each mass eigenstate. 
While such an approach might be appropriate for counting fringe
shifts in interference experiments \cite{stodolsky}, 
its application in Ref. \cite{BHM} to the calculation 
of the neutrino oscillation phase 
involves the assumption that the mass eigenstates are 
emitted at different times. The physical validity of this approach
is suspect.
In addition, it does not allow for 
possible gravitational
effects on the spin of the neutrinos \cite{PRW} (see Sec. III).
Furthermore, Refs. \cite{ahluwalia,stodolsky} separate out a 
``gravitational contribution'' to this phase. Such a separation 
is only possible for weak fields, making the 
``gravitationally induced phase''
a concept of limited utility. Related to this, some authors
have used methods that appear to mix flat- and curved-spacetime
thinking \cite{ahluwalia,grossman}. 
These particular treatments can leave the reader confused about the
precise meaning of quantities such as the ``energy,'' and the
nature of the coordinates 
(i.e., Do they reflect proper time and distance?). Covariant
calculations do not suffer from these difficulties of
interpretation, and are thus
preferable. 

In Sec. II we formulate a standard treatment of neutrino oscillations
in a more geometric framework. In Sec. III we generalize our
treatment to curved spacetime, with calculations in Schwarzschild
geometry in Sec. IV (vacuum oscillations) and Sec. V 
(Mikheyev-Smirnov-Wolfenstein (MSW) effect). 
Conclusions are given in Sec. VI.
We set $G = \hbar = c = 1$ throughout this paper.

\section{Simple geometric treatment of neutrino oscillations: 
	flat spacetime}

In this section we briefly review a simple, standard treatment of
neutrino oscillations \cite{kaiser}, and then present a geometric
version.

In a standard treatment, the neutrino state is written
\begin{equation}
|\Psi_{\alpha}(x,t) \rangle = \sum_j U_{\alpha j} \exp \left[ -i
	(Et - P_j x) \right] | \nu_j \rangle.  \label{basic}
\end{equation}
Here flavor (mass) indices are in greek (latin) letters. The matrix
elements $U_{\alpha j}$ comprise the transformation between the
flavor and mass bases. The subscript $\alpha$ on the left hand
side indicates that the neutrino was in flavor state $\alpha$ at
the initial position $x=0$ and time $t=0$. 
The mass eigenstates are taken to be
energy eigenstates with a common energy $E$;\footnote{For 
	discussion on whether neutrino
	mass eigenstates should be considered  
	momentum eigenstates, energy eigenstates,
	or neither, see for example Refs. 
	\cite{grossman,goldman}
	and references therein. These technicalities are
	not crucial in the present context.} 
the three-momenta of the
mass eigenstates are then 
\begin{equation}
P_j = \sqrt{E^2 - m_j^2} \approx E - {m_j^2 \over 2E},
\end{equation}
where $m_i$ is the rest mass corresponding to
mass eigenstate $| \nu_i \rangle$. 
To compute the oscillation probability at position $x$, a massless
neutrino trajectory is assumed, i.e. $x = t$:
\begin{equation}
|\Psi_{\alpha}(x,x) \rangle = \sum_j U_{\alpha j} \exp \left[ -i
	\left(m_j^2 \over 2E\right) x \right] | \nu_j
	 \rangle. \label{standard}
\end{equation}
This state is then used to compute the oscillation amplitude.
We note that the assumption of a null trajectory is 
necessary for the observation of oscillations; if the mass eigenstates
could be measured at different positions (or times), the
interference pattern would be destroyed.\footnote{One 
	might argue that to account for observation of the
	interference pattern, an {\em average} (rather than null)
	trajectory should be employed; see Ref. \cite{grossman}.
	In this case we would have $x = (\overline{P}/E) t$,
	where $\overline{P}$ is the average momentum of 
	the mass eigenstates having a common energy $E$.
	However, this reasonable argument has no 
	consequences for our study, since it merely introduces an
	overall phase common to all eigenstates. 
	We will here follow (perhaps unfortunate) tradition
	and employ null trajectories throughout this paper.}

While the form of Eq. (\ref{basic}) may not openly 
suggest it, the neutrino state at a given point in 
spacetime is frame-invariant. The quantities on the right hand side
of Eq. (\ref{basic})---the transformation coefficients between
flavor and mass bases, the phase, the mass eigenstates---all are
frame independent quantities. Since the connection between quantities
in flat and curved spacetime is most apparent when expressions
are written in a manifestly covariant manner, we introduce a
generalized form of Eq. (\ref{basic}):
\begin{equation}
|\Psi_{\alpha}(\lambda) \rangle = \sum_j U_{\alpha j} \exp \left( i
	\int_{\lambda_0}^{\lambda} \vec{P} \cdot \vec{p}_{\rm null} \, 
	d\lambda \right) | \nu_j \rangle.   \label{geom}
\end{equation}
In this expression, $\vec{P}$ is the four-momentum operator that 
generates spacetime translation of the mass eigenstates. 
The quantity $\vec{p}_{\rm null} = d\vec{x} /d\lambda$ 
is the (null) tangent vector to the neutrino's worldline
$\vec{x}(\lambda) = 
\left[t(\lambda),x(\lambda),y(\lambda),z(\lambda)\right]$; 
$\lambda$ is an
affine parameter of the worldline. 

We now show that Eq. (\ref{geom}) is equivalent to Eq. (\ref{basic}),
by simplifying Eq. (\ref{geom}) for neutrino propagation in the 
$x$ direction. Let $i\Omega$ denote the argument of the exponential
in Eq. (\ref{geom}). With $\vec{P} = (E, P^x, 0,0)$, and employing
the metric $\eta_{\mu\nu} = \rm{diag}[-1,1,1,1]$,
we have 
\begin{equation}
\Omega = -\int_{\lambda_0}^{\lambda} \left( E {dt\over d\lambda} 
	- P_x {dx \over d\lambda} \right) \, d\lambda,\label{om}
\end{equation}
where 
\begin{equation}
P_x =\eta_{x\mu} P^{\mu} \simeq E - {M^2 \over 2E},  \label{momop}
\end{equation}
and $M$ is the mass operator. After the mass operator has done
its work, we have
\begin{equation}
\omega_j = -\int_{\lambda_0}^{\lambda} \left[ E {(dt/d\lambda) \over
	(dx/d\lambda)} - \left(E - {m_j^2 \over 2E} \right) \right]
	\, {dx \over d\lambda} \, d\lambda, \label{almost}
\end{equation}
where $\omega_j$ is the phase of the $j$-th mass eigenstate.
Since
\begin{equation}
{(dt/d\lambda) \over (dx/d\lambda)} = {p^t_{\rm null} \over
	 p^x_{\rm null}}
	= 1,
\end{equation}
Eq. (\ref{almost}) reduces to
\begin{equation}
\omega_j = -\int_{x_0}^x {m_j^2 \over 2E} \, dx = 
	-{m_j^2 \over 2E} (x - x_0).
\end{equation}
This phase agrees with that in Eq. (\ref{standard}) (in which
$x_0=0$), suggesting that
the neutrino state as written in Eq. (\ref{geom}) is suitable for 
calculating the vacuum oscillation amplitude. Furthermore, 
the form of Eq. (\ref{geom}) suggests a straightforward
generalization to curved spacetime.

We now review how contributions to effective neutrino mass
arising from neutrino forward scattering off background matter 
can be included in the
above formalism. These effects are important because they can
give rise to, for example, the MSW effect.
Our treatment is essentially that found in Ref. \cite{fuller}. 

As an example, we take neutrino propagation through 
an electron background. In this case the Dirac equation 
can be cast in the form
\begin{equation}
\left[ \gamma^{\mu} (\partial_{\mu} + i A_{f\mu}{\cal P}_L) + M_f 
	\right]
	\psi_f = 0.  \label{dirac}
\end{equation}
(See Ref. \cite{weinberg} for the convention for the Dirac matrices
$\gamma^{\mu}$ that we employ.) 
Here $\psi_f$ is a column vector of spinors of different neutrino
flavors, and $M_f$ is the vacuum mass matrix in the flavor basis:
\begin{equation}
M_f^2 = U \pmatrix{m_1^2 &0 \cr 0& m_2^2} U^{\dagger},
\end{equation}
with
\begin{equation}
U = \pmatrix{\cos \theta & \sin \theta \cr -\sin \theta
	& \cos \theta}.
\end{equation}
The vector $A_f^{\mu}$  is the flavor-basis effective
potential matrix for interaction with the electron background:
\begin{equation}
A_f^{\mu} = \pmatrix{ -\sqrt{2} G_F N_e^{\mu} & 0 \cr 0 & 0}.
\end{equation}
In this expression, $G_F$ is the Fermi constant, and
$N_e^{\mu} = n_e u^{\mu}$ is the number current of the electron
fluid; $n_e$ is the electron density in the fluid rest frame, and
$u^{\mu}$ is the fluid's 4-velocity. ${\cal P}_L$ is the left-handed 
projection operator. (An index labeling the neutrino's helicity
must now be included in the eigenkets of Eq. (\ref{geom})). 
The form of Eq. (\ref{dirac}) suggests that
the mass shell relation read
\begin{equation}
(P^{\mu} + A_f^{\mu}{\cal P}_L)(P_{\mu} + A_{f\mu}{\cal P}_L) = 
-M_f^2.
\end{equation}
This expression can be derived by iteration of the Dirac 
equation, with the assumption
that the neutrino momentum is much larger than the inverse
scale height of the background matter.
Assuming that the electron background is at rest with
respect to the oscillation experiment, keeping only terms to
first order in $G_F$, and writing $\vec{P} = (E,P^x,0,0)$, we 
find
\begin{equation}
P_x \simeq E  - {1 \over 2E}\left(M_f^2 - V_f\right),\label{effmass}
\end{equation}
with 
\begin{equation}
V_f = \pmatrix{ -2\sqrt{2} G_F E n_e{\cal P}_L & 0 \cr 0 & 0}.
\end{equation}
The three-momentum operator $P_x$ now includes effective mass 
contributions from background matter, and can be used in Eq. 
(\ref{om}).

\section{Simple geometric treatment of neutrino oscillations: 
	curved spacetime}

In applying Eq. (\ref{geom}) in curved spacetime, we see that
evaluation of the argument in the exponential can become more
involved. This complication results from the dependence of the
metric on position. Additionally, 
we may worry about another gravitational effect: since
gravitational fields can cause gyroscopes to precess, perhaps
gravitational fields can also cause neutrino spin flips 
\cite{PRW}.  

How can effects on spin be incorporated into Eq. (\ref{geom})? 
Gravitational effects on the spin arise through the ``spin 
connection'' $\Gamma_{\mu}$ appearing in the Dirac equation
in curved spacetime \cite{weinberg2} (we here ignore background
matter effects):
\begin{equation}
\left[ \gamma^{a} e^{\mu}_a (\partial_{\mu} + \Gamma_{\mu}) 
	+ M 
	\right]
	\psi = 0.  \label{dirac2}
\end{equation}
In this equation and in the rest of this section, greek indices
refer to general curvilinear coordinates, while the latin indices
$a,b,c,d$ refer to locally inertial (Minkowski) 
coordinates. The tetrads $e^{\mu}_a$
connect these sets of coordinates.
The explicit expression for $\Gamma_{\mu}$ is
\begin{equation}
\Gamma_{\mu} = {1 \over 8}[\gamma^b, \gamma^c] e^{\nu}_b
	e_{c \nu;\mu}.
\end{equation}
Effects on spin can be incorporated into the three-momentum
operator (like Eq. (\ref{momop})) in an analogous manner to
background matter effects.

We must first simplify the Dirac matrix product in the spin
connection term. It can be shown that
\begin{equation}
\gamma^a [\gamma^b, \gamma^c] = 2 \eta^{ab} \gamma^c -
	2 \eta^{ac} \gamma^b - 2i \epsilon^{dabc} \gamma_5 
	\gamma_d,  \label{gammas}
\end{equation}
where $\eta^{ab}$ is the metric in flat space and 
$\epsilon^{abcd}$ is the (flat space) totally antisymmetric tensor,
with $\epsilon^{0123}= +1$. With Eq. (\ref{gammas}), the
non-vanishing contribution from the spin connection is
\begin{equation}
\gamma^a e_a^{\mu} \Gamma_{\mu} = \gamma^a e_a^{\mu}
	\left\{  i
	A_{G\mu} \left[-(-g)^{-1/2}{\gamma_5 \over 2}\right] \right\},
	\label{gravpot}
\end{equation}
where 
\begin{equation}
A^{\mu}_G \equiv {1 \over 4} (-g)^{1/2} e^{\mu}_a \epsilon^{abcd}
	(e_{b\nu,\sigma} - e_{b\sigma,\nu}) e^{\nu}_c
	e^{\sigma}_d.
\end{equation}
In these equations, $(-g)^{1/2} = [\rm det(g_{\mu\nu})]^{1/2}$,
where $g_{\mu\nu}$ is the metric of curved spacetime.
The expression in Eq. (\ref{gravpot}) treats left- and right-handed
states differently. In order to group it with terms arising
from matter effects, we can without physical consequence
add a term proportional to the identity to obtain
\begin{equation}
\gamma^a e_a^{\mu} \Gamma_{\mu} = \gamma^a e_a^{\mu}
	(i A_{G\mu} {\cal P}_L).
\end{equation}
Proceeding as in the discussion of matter effects in the last
section, the 3-momentum operator used in neutrino 
oscillation calculations can be computed from the mass
shell condition
\begin{equation}
(P_{\mu} + A_{G\mu}{\cal P}_L)(P^{\mu} + A_G^{\mu}{\cal P}_L) = -M^2,
\end{equation}
where we have not included background matter effects.

An important point is that the gravitational contribution
$A_G^{\mu}$ is proportional to the identity matrix in
flavor space, and diagonal in spin space. It cannot induce
spin flips on its own.
Therefore, it will not have any observable effects 
unless there are other off-diagonal terms in spin space
(e.g., from the interaction of a neutrino magnetic moment
with a magnetic field) \cite{PRW}.

Another complication in applying Eq. (\ref{geom})
in curved spacetime is related to the nature of the
neutrino trajectories. In flat spacetime, the neutrino
trajectories are straight lines. The propagation can
be taken to be in one spatial dimension, and the variable
of integration becomes the spatial variable corresponding
to that direction of propagation, as in Sec. II. However,
the neutrino trajectories in curved spacetime are typically
parametrized curves involving more than one spatial variable:
$\vec{x}(\lambda) = 
\left[x^0(\lambda),x^1(\lambda),x^2(\lambda),x^3(\lambda)\right]$.

For general trajectories it therefore may be convenient to leave the
affine parameter $\lambda$ as the variable of integration, as in
Eq. (\ref{geom}). The tangent vector to the null worldline, 
$\vec{p} \equiv \vec{p}_{\rm null} = d\vec{x}/d\lambda$, 
can be found from the
geodesic equation or Hamilton-Jacobi equation. The 
four-momentum operator $\vec{P}$ can be constructed as follows:
(1) take the neutrinos to be energy eigenstates, and set 
$P^0=p^0$; (2) demand that the three-momenta of $\vec{P}$ and
$\vec{p}$ be parallel, i.e. $P^i = p^i (1-\epsilon)$ with
$i=1,2,3$; and (3) 
$(P_{\mu}+A_{\mu}{\cal P}_L)(P^{\mu}+A^{\mu}{\cal P}_L) = -M^2$,
with $A^{\mu}$ now representing both matter and ``spin connection''
contributions. For relativistic neutrinos $(\epsilon \ll 1)$,
ignoring terms of ${\cal O}(A^2)$ and ${\cal O}(AM^2)$, 
and remembering
that $\vec{p}$ is a null vector, we find
\begin{equation}
(g_{0i}p^0 p^i + g_{ij}p^i p^j)\epsilon = {M^2 \over 2} + \vec{p}
	\cdot \vec{A}{\cal P}_L.
\end{equation}
(Here the indices $i,j$ refer to the spacelike general
curvilinear coordinates, not locally inertial coordinates.)
From this it follows that the quantity $\vec{P} \cdot \vec{p}$
appearing in Eq. (\ref{geom}) is simply
\begin{equation}
\vec{P} \cdot \vec{p} = -\left({M^2 \over 2} + \vec{p} \cdot 
	\vec{A}{\cal P}_L
		\right). \label{operator}
\end{equation}
It is convenient to define a column vector of flavor amplitudes. 
For example, for mixing between $\nu_e$ and $\nu_{\tau}$,
\begin{equation}
\chi(\lambda) \equiv \pmatrix{\langle \nu_e | \Psi(\lambda) 
		\rangle \cr
		\langle \nu_{\tau} | \Psi(\lambda) \rangle}.
\end{equation}
Eq. (\ref{geom}) can be written as a differential equation for
$\chi(\lambda)$, 
\begin{equation}
i {d\chi \over d\lambda} = \left({M_f^2 \over 2} + \vec{p} 
	\cdot \vec{A_f}{\cal P}_L	\right) \chi, \label{affdiff}
\end{equation}
where the subscript $f$ denotes `flavor basis.' Eq. (\ref{affdiff})
can be integrated (numerically if necessary) to yield the
neutrino flavor evolution. A similar equation was obtained in
Ref. \cite{PRW} by more formal methods.

\section{Neutrino oscillations in Schwarzschild spacetime:
	vacuum oscillations}

While Eq. (\ref{affdiff}) may be useful for calculating the
neutrino flavor evolution for general neutrino trajectories  
in general spacetimes, it does not yield a great deal of
physical insight. In this and the following section we do
example calculations in Schwarzschild geometry. 
This example geometry is simple
enough that the oscillation formulae can be cast in a form 
that resembles the flat space case. 

In this section we contrast radially propagating neutrinos
with azimuthally propagating neutrinos in order to demonstrate
how gravity affects vacuum neutrino oscillations. 
The geometry in a spherically symmetric, static spacetime
can be globally represented by the Schwarzschild coordinate system
$\{x^{\mu}\}\Rightarrow (t,r,\theta,\phi)$. We can take
the Schwarzschild line element, which serves to
define these coordinates, as
\begin{eqnarray}
ds^2 &=& g_{\mu\nu}dx^{\mu}dx^{\nu} \nonumber \\ 
	&=& -e^{2\Phi(r)} dt^2 + e^{2\Lambda(r)} dr^2 \nonumber \\
	& &+ r^2 d\theta^2 + r^2 \sin ^2\theta d\phi ^2.  
\end{eqnarray}
Using the tetrads
\begin{equation}
e^{\mu}_a = {\rm diag}\left[ e^{-\Phi (r)},
	e^{-\Lambda (r)}, {1\over r}, 
	{1 \over r \sin \theta} \right],	
\end{equation}
direct calculation yields $A_G^{\mu}=0$. This is perhaps 
expected from spherical symmetry, and is
in agreement
with Ref. \cite{PRW}, in which terms arising from the spin
connection vanish in the Schwarzschild geometry.
Since the components of the metric are independent of the timelike 
coordinate
$t$, there is a conserved quantity, the timelike covariant
momentum component $P_t \equiv -E_*$. (The relation between
covariant and contravariant components is $P_{\mu} = 
g_{\mu\nu} P^{\nu}$.)
We take the neutrino
states to be eigenstates of this quantity. 

Denoting a differential proper distance at constant $t$ by
$d\ell$, we can write
\begin{eqnarray}
d\lambda &=& d\ell \left(g_{ij} {dx^i\over d\lambda}
	{dx^j\over d\lambda}\right)^{-1/2} \nonumber \\
	&=& d\ell \left[-g_{00} 
	\left({dx^0\over d\lambda}\right)^2\right]^{-1/2}, 
\end{eqnarray}
where we have used the facts that the neutrino trajectory is null
and that the Schwarzschild metric does not mix time and space
components. Using this expression for $d\lambda$ and Eq. 
(\ref{operator}), we obtain in vacuum
\begin{equation}
\Omega = \int_{\lambda_0}^{\lambda} \vec{P} \cdot 
	\vec{p}_{\rm null} \, 
	d\lambda = -\int_{\ell_0}^{\ell} {M^2 \over {2E_l}}\, d\ell.
	\label{schwom}
\end{equation}
In this equation, $E_l = E_*e^{-\Phi (r)}$  is the energy measured
by a locally inertial observer 
momentarily at rest in the Schwarzschild spacetime
(and presumably at rest with respect to the 
``oscillation experiment''). Therefore,
the integrand in Eq. (\ref{schwom}) is formally the same as
the corresponding integrand in flat space. 

For radial propagation, $d\ell = e^{\Lambda (r)} dr$ 
is a differential element of
proper distance for constant $t,\ \theta,\ \phi$, and so
\begin{equation}
\Omega = -\int_{\ell_0}^{\ell} {M^2 \over {2E_l}}\, d\ell
	= -\int_{r_0}^r {M^2 \over {2E_*e^{-\Phi (r)}} } \, 
	e^{\Lambda (r)} dr. \label{omr}
\end{equation}
We see that unlike
the flat space case, the integral in terms of proper distance
is not trivial, due to
the gravitational redshift of the ``local energy'' $E_l$ and the
radial dependence of $d\ell$. In this manner, spacetime curvature 
(gravity) makes its
impact on the oscillation amplitude.

Of course, in vacuum above a spherical, static source of 
gravitational mass ${\cal M}$, we have
\begin{eqnarray}
e^{2\Phi(r)} &=& \left(1 - { r_s \over r} \right), \\
e^{2\Lambda(r)} &=&  \left(1 - { r_s \over r} \right)^{-1},
\end{eqnarray}
where the Schwarzschild radius is $r_s \equiv 2{\cal M}$. 
Then Eq. (\ref{omr}) is
trivially integrated:
\begin{equation}
\Omega= -{M^2 \over 2E_*}
	(r - r_0). 
\end{equation}
Again, the coordinate difference $(r-r_0)$ 
does {\em not} reflect a proper distance. (The {\em proper} distance
$\ell$ corresponding to the {\em coordinate} difference $(r-r_0)$ is
$\ell = \int_{r_0}^r \sqrt{g_{rr}}\,dr = 
\int_{r_0}^r e^{\Lambda (r)}\, dr$.)
Likewise,
$E_*$ does {\em not} represent the neutrino 
energy measured by a locally
inertial observer at
rest at finite radius, but rather the energy of the neutrino 
measured by such an observer
at rest at infinity. It is generally not possible to extract a 
separate ``gravitational phase'' from this expression; nevertheless,
it is clear that gravity has an effect on the oscillations of  
radially propagating neutrinos. In the weak field limit one could
define a ``gravitational phase,'' however.

In the Schwarzschild spacetime there are circular orbits of radius
$r = R \equiv 3{\cal M}$ for massless particles. Consideration of 
neutrino oscillations in such an orbit (which 
we take to be in the plane defined by
$\sin \theta = 1$) can provide insight into gravitational effects
in the azimuthal direction.

For azimuthal propagation in this orbit, $d\ell = R d\phi$, and
the local energy $E_l = E_* \left(1 - r_s/R \right)^{-1/2}$ is
constant along the neutrino trajectory, in contrast to the case of
radial propagation. From Eq. (\ref{schwom}) we find
\begin{equation}
\Omega = -{M^2 \over 2E_l} R(\phi - \phi_0). \label{azphase}
\end{equation}
This expression involves local energy and proper distance,
making it precisely the same as the corresponding flat space 
expression.
Gravity has no effect here.\footnote{We note 
	that a ``gravitational phase'' for motion 
	transverse to the radial 
	direction is given in Ref. \cite{ahluwalia}, 
	in contrast with our result.
	The discrepancy arises because those authors
	attempt to use a gravitational potential energy
	instead of employing covariant methods.}

\section{Neutrino oscillations in Schwarzschild spacetime:
	MSW effect}

In this section we study an example of gravitational 
effects on MSW resonant neutrino transformations 
\cite{msw,haxton}. 
In particular, 
in Schwarzschild spacetime 
we find the resonance position and calculate the 
adiabaticity parameter for a radially propagating 
two-flavor neutrino system
in an electron background with monotonically decreasing density
profile.

It will here be convenient to write the neutrino evolution
equation in terms of column vectors of flavor amplitudes:
\begin{equation}
\chi(r) \equiv \pmatrix{\langle \nu_e | \Psi(r) \rangle \cr
		\langle \nu_{\tau} | \Psi(r) \rangle}.
\end{equation}
For radial propagation we obtain
\begin{equation}
\chi(r)\! =\! \exp \! \left[-i\!\! 
	\int_{r_0}^r \!{1\over {2 E_*}} \!\left[e^{\Phi (r)} M_f^2 
	+  V_f(r) \right] e^{\Lambda (r)} dr \right]\!
	\chi(r_0). \label{matterint}
\end{equation}
In this equation, $M_f^2$ is the vacuum mass matrix in the flavor
basis.
The contribution from the background matter is
\begin{equation}
V_f(r) = \pmatrix{v(r) & 0 \cr 0 & 0},
\end{equation}
with $v(r) = 2\sqrt{2} G_F E_* n_e$ and $n_e = u_{\mu}
N_e^{\mu}$ is the locally measured electron density. 

Eq. (\ref{matterint}) can also be written as a Sch\"odinger-like
equation,
\begin{equation}
i{d\chi(r) \over dr}= {\tilde M_f^2 \over {2 E_*}} \chi(r),
\label{matterdiff}
\end{equation}
where the effective mass matrix in the flavor basis is
\begin{equation}
\tilde M_f^2 = e^{[\Lambda (r) +
	\Phi (r)]} M_f^2 + e^{\Lambda (r)} V_f(r).
\end{equation}
The mixing angle in matter, $\tilde \theta$, is defined in terms of
the diagonalization of $\tilde M_f^2$:
\begin{eqnarray}
\tilde M^2 &=& \tilde U^{\dagger} \tilde M_f^2 \tilde U = 
	\pmatrix{\tilde m_1^2 & 0 \cr 0 & 
	\tilde m_2^2}, \\ \nonumber \\
\tilde U &=& \pmatrix{ \cos \tilde\theta & \sin \tilde \theta \cr
		-\sin \tilde \theta & \cos \tilde \theta}.
\end{eqnarray}
The difference of the squares of the neutrino mass eigenvalues in
matter, $\tilde \Delta \equiv \tilde m_2^2 - \tilde m_1^2$, is
given in terms of $\Delta \equiv m^2_2 - m_1^2$ as
\begin{equation}
\tilde \Delta = e^{\Lambda} \left[\left(v - e^{\Phi}\Delta
	\cos 2\theta\right)^2 + \left(e^{\Phi}\Delta \sin
	2\theta\right)^2\right]^{1/2}.
\end{equation}
The mixing angle in matter is given by
\begin{equation}
\tan 2\tilde \theta = {e^{\Phi} \Delta \sin 2\theta \over
	{\left(-v + e^{\Phi} \Delta \cos 2\theta \right)}}.
\end{equation} 

The ``resonance'' occurs for $\sin^2 2\tilde \theta = 1$, where
a mass level crossing occurs and $\tilde \Delta$
is a minimum. The resonance condition is $v(r) = e^{\Phi (r)}
\Delta \cos 2\theta$, or 
\begin{equation}
\sqrt{2} G_F n_e = {\Delta \over {2 e^{-\Phi} E_*}} \cos 2\theta
	= {\Delta \over {2 E_l}} \cos 2\theta,
\end{equation}
where $E_l = -\vec p \cdot \vec u = E_*e^{-\Phi}$
is the redshifting ``local energy'' introduced
in the last section. Unlike flat space, the resonance condition
is here determined in part by an energy which may be redshifted
from the energy the neutrino was born with at the production site.

After dropping a term proportional to the identity matrix
that yields an overall phase, Eq. (\ref{matterdiff}) can be
written in the basis of instantaneous mass eigenstates as
\begin{equation}
i{d\tilde\chi(r) \over dr} = \pmatrix{
	-\tilde\Delta / {4E_*} &
	-i d\tilde \theta / dr \cr i d\tilde \theta /dr & 
	\tilde\Delta / {4E_*}} \tilde \chi(r), \label{addiff}
\end{equation}
where
\begin{equation}
\tilde \chi(r) = \pmatrix{\langle \tilde \nu_1 | \Psi(r) \rangle \cr
		\langle \tilde \nu_2 | \Psi(r) \rangle},
\end{equation}
and $|\tilde \nu_1 \rangle$ and $|\tilde \nu_2 \rangle$ are the
instantaneous mass eigenstates. 
The adiabaticity parameter $\gamma (r)$, which compares the relative 
magnitudes of the diagonal and off-diagonal terms in Eq. 
(\ref{addiff}), is defined to be 
\begin{equation}
\gamma (r) = {\tilde \Delta \over {4E_* |d\tilde\theta / dr|}}.
\end{equation}
For $\gamma \gg 1$, the neutrino evolution can be approximated
by a constant superposition of slowly varying instantaneous
mass eigenstates, except for a small probability 
$\exp\left[-{\pi \over 2} \gamma (r_{\rm res})\right]$
for one
mass eigenstate to jump to the other at resonance, 
where $r_{\rm res}$ is the position of the resonance
\cite{haxton}.
We find
\begin{eqnarray}
\gamma (r_{\rm res}) &=& {\Delta \sin ^2 2\theta \over
	e^{-\Phi} E_* \cos 2 \theta} \left| e^{-\Lambda}
	{d \over dr}\ln \left(e^{-\Phi} n_e\right)  \right|^{-1}
	\nonumber \\ \nonumber \\
	&=&  {\Delta \sin ^2 2\theta \over
	E_l \cos 2 \theta} \left| e^{-\Lambda}
	{d\over dr} \ln \left(-\vec p \cdot \vec A\right) 
	 \right|^{-1}, \label{ad}
\end{eqnarray}
where $\vec A$ is the matter potential four-vector introduced
in Sec. I. Thus the adiabaticity of the evolving neutrino
(the degree to which the ``jump probability'' is unimportant) 
is affected by the spatial dependence of the metric, which 
appears for example in the determination of the local energy.

\section{Conclusion}

We have developed a simple formalism for treating 
neutrino oscillations
in curved spacetime. This formalism can accomodate 
matter effects and
gravitational contributions to neutrino spin precession 
in the presence
of a magnetic field.

We have done explicit calculations in 
Schwarzschild spacetime (without background matter
effects). Our simple formalism has verified the result of 
Ref. \cite{PRW} that
gravitational contributions to spin precession
vanish in spherically symmetric, static Schwarzschild spacetime.
However, we have found that the oscillation formulae for 
radially propagating neutrinos are
altered by gravity. This alteration results from the metrical
properties of curved spacetime. The basis of the effects we found
are closely related to the gravitational redshift, in the case
of both vacuum oscillations and the MSW effect.
In contrast, azimuthally
propagating neutrinos show 
no alterations to their oscillation formulae in the spherically
symmetric, static case.

In applications where strong gravitational effects on neutrino 
oscillations are of possible interest (e.g., supernovae),
matter effects will generally make vacuum oscillations 
(and therefore 
gravitational effects on the vacuum oscillation phase) 
unimportant. However,
gravitational effects on the resonance position and adiabaticity
of the MSW effect are of potential interest.

For example, the requirement for successful $r$-process 
nucleosynthesis in a neutron-rich post--core-bounce
supernova environment 
has been used to delineate values of neutrino mass difference
and mixing angle which favor/disfavor heavy
element production \cite{qian93}. 
These limits arise because the 
$\nu_{\mu}$ and  $\nu_{\tau}$ neutrinos emitted from the 
supernova have a higher average energy than the emitted
$\nu_e$ neutrinos. Therefore, an MSW resonant transformation
of $\nu_{\mu}$ or $\nu_{\tau}$ neutrinos provides a population
of higher energy $\nu_e$ neutrinos that tend to drive the
material outside the nascent neutron star toward less
neutron-rich conditions. In order to preclude $r$-process
nucleosynthesis, the MSW transformation must be sufficiently
adiabatic (conversion efficiency $\sim 30$\%), and must occur
before the radius where the neutron-to-proton ratio freezes
out (``weak freeze-out radius''). As we saw in Sec. V, 
gravitation affects both the adiabaticity parameter and
the position of the resonance. 

Current supernova models indicate that the 
general relativistic effects we consider here are 
probably not very important. However,
the equation of state of nuclear matter is not well
understood, and it may be that  during the
time frame of interest for nucleosynthesis
some proto-neutron stars
may become very relativistic \cite{bb94}. 
If this turns out to be the case, we may
hazard the following conjectures regarding gravitational
effects on limits on neutrino mass difference and mixing
angle from $r$-process considerations. The MSW transformation
may become more adiabatic because of the redshifting energy
appearing in the denominator of the adiabaticity parameter.
This would extend the limit on neutrino mixing to exclude
smaller values of the vacuum mixing angle. On the other
hand, the smallest neutrino mass difference excluded by
$r$-process considerations is, roughly, that mass 
difference for which the resonance position of an 
average energy $\mu$ (or $\tau$) neutrino coincides with
the weak freeze-out radius. The redshifting energy 
appearing in the 
resonance condition will tend to pull the resonance position
closer to the neutron star. However, the weak freeze-out
radius is determined by the competition between the
weak interaction rates and the expansion rate of the material
outside the the neutron star, and the weak interaction rates
are proportional to the {\em square} of the redshifting
neutrino energy. Therefore, for a given mass difference,
the redshift will reduce the separation of the 
resonance position and the weak freeze-out radius. This
would {\em weaken} the mass difference boundary of the
excluded region. Of course, these conjectures are
preliminary, as they are based principally on redshift effects.
Other effects of a supernova core of sufficiently small
radius for gravitational effects to become interesting
may be relevant. Such effects might include 
changes in the density scale height 
and expansion rate of the background matter in the supernova
envelope, and alteration of the neutrino spectrum. In fact, if
the $\nu_e$ and ${\overline \nu}_e$ are significantly redshifted,
then $r$-process nucleosynthesis may be precluded anyway
\cite{fuller96}.

\section*{Acknowledgements}
We are grateful for helpful conversations with A. B. Balantekin, 
T. Goldman, H. J. Lipkin,
and Y.-Z. Qian. This work was supported 
by grants NSF PHY95-03384 and NASA NAG5-3062
at UCSD.


\begin{references}

\bibitem{colella}R. Collela, A. W. Overhauser, and S. A. Werner,
	Phys. Rev. Lett. {\bf 34}, 1472 (1975).

\bibitem{ahluwalia}D. V. Ahluwalia and C. Burgard,
	LA-UR-96-862, gr-qc/9603008 (1996);
	D. V. Ahluwalia and C. Burgard, LA-UR-96-2031, gr-qc/9606031
	(1996).

\bibitem{grossman}Y. Grossman and H. J. Lipkin, WIS-96/27/Jun-PH,
	TAUP 2346-96, hep-ph/9607201 (1996).

\bibitem{BHM}T. Bhattacharya, S. Habib, and E. Mottola, Los Alamos
	Preprint, gr-qc/9605074 (1996).

\bibitem{PRW}D. P\'{\i}riz, M. Roy, and J. Wudka, 
	Phys. Rev. D {\bf 54},
	1587 (1996).

\bibitem{fuller96}G. M. Fuller and Y.-Z. Qian, Nucl. Phys. A
	{\bf 606}, 167 (1996).

\bibitem{stodolsky}L. Stodolsky, Gen. Rel. and Grav., {\bf 11}, 391 
	(1979).

\bibitem{kaiser}See e.g. B. Kayser, Phys. Rev. D {\bf 24}, 110 (1981). 

\bibitem{goldman}T. Goldman, LA-UR-96-1349, hep-ph/9604357 (1996).

	
\bibitem{fuller}G. M. Fuller, R. W. Mayle, J. R. Wilson, and
	D. N. Schramm, Astrophys. J. {\bf 322}, 795 (1987).

\bibitem{weinberg}S. W. Weinberg,
	{\em The Quantum Theory of Fields} (Cambridge UP,
	Cambridge, 1995), Sec. 5.4.

\bibitem{weinberg2}See e.g. S. W. Weinberg,
	{\em Gravitation and Cosmology} (Wiley, New York,
	1972), Sec. 12.5; D. R. Brill and J. A. Wheeler,
	Rev. Mod. Phys. {\bf 29}, 465.

\bibitem{msw}L. Wolfenstein, Phys. Rev. D {\bf 17}, 
	2369 (1978); S. Mikheyev and
	A. Yu. Smirnov, Nuovo Cimento Soc. Ital. Fis. C 
	{\bf 9}, 17 (1986).

\bibitem{haxton}W. C. Haxton, Phys. Rev. D {\bf 35}, 2352 
	(1987).

\bibitem{qian93}G. M. Fuller, Phys. Rep. {\bf 227}, 149 (1993);
	Y.-Z. Qian, G. M. Fuller, G. J. Mathews,
	R. W. Mayle, J. R. Wilson, and S. E. Woosley,
	Phys. Rev. Lett. {\bf 71}, 1965 (1993);
	Y.-Z. Qian and G. M. Fuller, Phys. Rev.
	D {\bf 51}, 1479 (1995).

\bibitem{bb94} G. E. Brown and H. A. Bethe, Astrophys. J., {\bf 
423}, 659 (1994).


\end{references}
\end{document}